\documentclass[a4paper,12pt]{article}
\pdfoutput=1
\usepackage{feynmp-auto,expdlist,contour}
\usepackage{amsmath, amsfonts}
\usepackage{graphicx,arydshln}
\usepackage{enumerate}
\usepackage{hyperref}
\usepackage{latexsym}
\usepackage{dsfont}
\usepackage{hepnicenames}
\usepackage{enumerate}
\usepackage{soul}
\usepackage[normalem]{ulem}
\usepackage{comment}

\newcommand{\footnoten}[1]{}

\usepackage[font={small},textfont={it}]{caption} 

\usepackage{units}

\usepackage{mathrsfs,graphicx,rotating,amsmath,amsfonts,mathtools,booktabs,amssymb,wasysym}
\usepackage{hyperref}\usepackage{slashed}
\usepackage[nosort]{cite}
\usepackage[table,xcdraw,dvipsnames]{xcolor}
\usepackage{bm}
\usepackage{multirow,multicol}
\hypersetup{colorlinks,bookmarksopen,bookmarksnumbered,
	linkcolor=blus,pdfstartview=FitH,urlcolor=rossos,citecolor=verde}
\allowdisplaybreaks

\newcommand{\myfootnote}[1]{}

\renewcommand{\[}{\left[}

\newcommand{\bP}{\bar{M}_{\rm Pl}}

\def\Lag{\mathscr{L}}

\newcommand{\mio}[1]{}

\newcommand{\med}[1]{\langle #1\rangle}

\def\bpm{\begin{pmatrix}}
	\def\epm{\end{pmatrix}}

\usepackage{mathrsfs}

\newcommand{\fig}[1]{~\ref{fig:#1}}
\newcommand{\sfrac}[2]{#1/#2}

\newcommand{\nnn}[1]{{\color{verdes}#1}}

\allowdisplaybreaks
\usepackage{multicol}
\usepackage{color}
\definecolor{rosso}{cmyk}{0,1,1,0.4}
\definecolor{rossos}{cmyk}{0,1,1,0.55}
\definecolor{rossoc}{cmyk}{0,1,1,0.2}
\definecolor{blu}{cmyk}{1,1,0,0.3}
\definecolor{blus}{cmyk}{1,1,0,0.6}
\definecolor{bluc}{cmyk}{1,1,0,0.1}
\definecolor{verde}{cmyk}{0.92,0,0.59,0.25}
\definecolor{verdec}{cmyk}{0.92,0,0.59,0.15}
\definecolor{verdes}{cmyk}{0.92,0,0.59,0.4}

\newcommand{\bp}{\bar{M}_{\rm Pl}}
\newcommand{\eq}[1]{~{\rm (\ref{eq:#1})}}

\newcommand{\GeV}{\,{\rm GeV}}

\def\circa#1{\,\raise.3ex\hbox{$#1$\kern-.75em\lower1ex\hbox{$\sim$}}\,}

\newcommand{\beq}{\begin{equation}}
\newcommand{\eeq}{\end{equation}}

\newcommand{\bea}{\begin{eqnarray}}
\newcommand{\eea}{\end{eqnarray}}
\newcommand{\be}{\begin{equation}}
\newcommand{\ee}{\end{equation}}
\font\tenrsfs=rsfs10 at 12pt
\font\sevenrsfs=rsfs7
\font\fiversfs=rsfs5
\newfam\rsfsfam
\textfont\rsfsfam=\tenrsfs
\scriptfont\rsfsfam=\sevenrsfs
\scriptscriptfont\rsfsfam=\fiversfs

\newsavebox\MBox

\renewenvironment{thebibliography}[1]
{\begin{multicols}{2}[\section*{\refname}]%
		\@mkboth{\MakeUppercase\refname}{\MakeUppercase\refname}%
		\list{\@biblabel{\@arabic\c@enumiv}}%
		{\settowidth\labelwidth{\@biblabel{#1}}%
			\leftmargin\labelwidth
			\advance\leftmargin\labelsep
			\@openbib@code
			\usecounter{enumiv}%
			\let\p@enumiv\@empty
			\renewcommand\theenumiv{\@arabic\c@enumiv}}%
		\sloppy
		\clubpenalty4000
		\@clubpenalty \clubpenalty
		\widowpenalty4000%
		\sfcode`\.\@m}
	{\def\@noitemerr
		{\@latex@warning{Empty `thebibliography' environment}}%
		\endlist\end{multicols}}

\renewcommand{\L}\Lag

\def\circa#1{\,\raise.3ex\hbox{$#1$\kern-.75em\lower1ex\hbox{$\sim$}}\,}
\makeatletter

\font\ital=cmu10

\def\hhref#1{\href{http://arxiv.org/abs/#1}{arXiv:#1}}
\usepackage{xstring}
\newcommand{\hhrefq}[1]{\IfSubStr{#1}{:}{\href{http://inspirehep.net/search?ln=en&ln=en&p=#1&of=hb&action_search=Search&sf=&so=d&rm=&rg=25&sc=0}{InSpire:#1}}{\hhref{#1}}}

\def\art{\@ifnextchar[{\eart}{\oart}}
\def\eart[#1]#2#3#4#5#6{{\rm #2}, {\em #3 \bf #4} {\rm (#6) #5} ({\em #1})}
\def\article{\@ifnextchar[{\earticle}{\oarticle}}
\def\oarticle#1#2#3#4#5#6{{\rm #1}, {\ital `#6'}, {\rm #2 #3 (#5) #4}}
\def\earticle[#1]#2#3#4#5#6#7{{\rm #2}, {\ital `#7'}, {\rm #3 #4 (#6) #5}  [\hhrefq{#1}]}
\def\hepart[#1]#2{{\rm #2, \sl#1}}
\def\heparticle[#1]#2#3{#2, {\ital `#3'} [\hhrefq{#1}]}
\newcommand{\doi}[1]{\href{http://dx.doi.org/#1}{[link]}}

\newcommand{\hhrefqq}[1]{\IfBeginWith{#1}{10.}{\href{https://doi.org/#1}{doi:#1}}{\hhrefq{#1}}}
\def\earticle[#1]#2#3#4#5#6#7{{\rm #2}, {\ital `#7'}, {\rm #3 #4 (#6) #5}  [\hhrefqq{#1}]}

\renewenvironment{thebibliography}[1]
{\begin{multicols}{2}[\section*{\refname}]%
		\@mkboth{\MakeUppercase\refname}{\MakeUppercase\refname}%
		\list{\@biblabel{\@arabic\c@enumiv}}%
		{\settowidth\labelwidth{\@biblabel{#1}}%
			\leftmargin\labelwidth
			\advance\leftmargin\labelsep
			\@openbib@code
			\usecounter{enumiv}%
			\let\p@enumiv\@empty
			\renewcommand\theenumiv{\@arabic\c@enumiv}}%
		\sloppy
		\clubpenalty4000
		\@clubpenalty \clubpenalty
		\widowpenalty4000%
		\sfcode`\.\@m}
	{\def\@noitemerr
		{\@latex@warning{Empty `thebibliography' environment}}%
		\endlist\end{multicols}}

\newcounter{alphaequation}[equation]
\def\thealphaequation{\theequation\hbox to
	0.6em{\hfil\alph{alphaequation}\hfil}}
\def\eqnsystem#1{
	\def\@eqnnum{{\rm (\thealphaequation)}}
	\def\@@eqncr{\let\@tempa\relax \ifcase\@eqcnt \def\@tempa{& & &} \or
		\def\@tempa{& &}\or \def\@tempa{&}\fi\@tempa
		\if@eqnsw\@eqnnum\refstepcounter{alphaequation}\fi
		\global\@eqnswtrue\global\@eqcnt=0\cr}
	\refstepcounter{equation} \let\@currentlabel\theequation \def\@tempb{#1}
	\ifx\@tempb\empty\else\label{#1}\fi
	\refstepcounter{alphaequation}
	\let\@currentlabel\thealphaequation
	\global\@eqnswtrue\global\@eqcnt=0 \tabskip\@centering\let\\=\@eqncr
	$$\halign to \displaywidth\bgroup \@eqnsel\hskip\@centering
	$\displaystyle\tabskip\z@{##}$&\global\@eqcnt\@ne
	\hskip2\arraycolsep\hfil${##}$\hfil& \global\@eqcnt\tw@\hskip2\arraycolsep
	$\displaystyle\tabskip\z@{##}$\hfil
	\tabskip\@centering&\llap{##}\tabskip\z@\cr}
\def\endeqnsystem{\@@eqncr\egroup$$\global\@ignoretrue} \makeatother

\definecolor{Gray}{gray}{0.95}

\def\bal#1\eal{\begin{align}#1\end{align}}

\newcommand{\TBH}{T_{\rm BH}}

\begin{document}
\begin{center}  \contourlength{0.2pt}
\resizebox{0.7\textwidth}{!}{\bf\color{rossos}\contour{red!20!black}{Black holes don't source}}\\[1ex]
\resizebox{0.7\textwidth}{!}{\bf\color{rossos}\contour{red!20!black}{fast Higgs vacuum decay}}\\[4ex]

{\bf\large Alessandro Strumia}\\[2ex]
{\it Dipartimento di Fisica, Universit\`a di Pisa, Italia}\\[3ex]

{\large\bf Abstract}\begin{quote}
We  argue
 that the rate of Standard Model vacuum or thermal decay seeded by primordial black holes is negligible (because non-perturbatively suppressed by the small quartic Higgs coupling) and independent of the non-minimal coupling of the Higgs to gravity.
\end{quote}
\end{center}
\tableofcontents

\section{Introduction}
Black holes and vacuum decay are two subtle quantum subjects,
and this is well reflected by the confusion in the literature about computing
the rate of vacuum decay seeded by black holes.

First attempts around 1990~\cite{Hiscock:1987hn,Arnold:1989cq,Berezin:1990qs,Samuel:1992wt}
reached the following understanding~\cite{Arnold:1989cq}:
the problem is difficult because it cannot be solved using the tool 
that allowed Coleman to compute the decay rate of Minkowski and of (anti-)de Sitter vacua:
analytic continuation to imaginary (Euclidean) time.
The reason is that the Euclidean Schwarzschild metric
describes the Hartle-Hawking vacuum~\cite{Hartle:1976tp}, where
a black hole of mass $M$ is in unstable static equilibrium with a thermal bath 
at the Hawking temperature $\TBH = 1/8\pi G M$ that fills the entire space-time.
We are instead interested in vacuum decay seeded by a black hole in empty space, 
a dynamical system that emits radiation at the Hawking temperature,
as described by the Unruh vacuum~\cite{Unruh:1976db,1309.0530}. 
In 1990 Arnold concluded `I do not know how to handle'~\cite{Arnold:1989cq} and
the issue remained quiescent getting partially forgotten.

\smallskip

Interest in vacuum decay seeded by black holes
resurfaced in the past decade~\cite{1401.0017,1501.04937,1503.07331,Iliesiu:2015lka,1601.02152,1606.04018,1704.05399,1706.01364,1706.04523,1708.02138,2005.12808,2105.09331,2111.08017,2209.00639}.
The measurement of the Higgs boson mass $M_h \approx 125\GeV$~\cite{1207.7235,1207.7214}
implied a possible instability in the effective Higgs potential~\cite{1112.3022,1307.3536}, given by
$V_{\rm eff}(h\gg M_h) \simeq \lambda(h) h^4/4$ in flat space.
The SM quartic Higgs coupling can turn negative, $\lambda(h)<0$, when extrapolated up to a critical field value, $h > h_{\rm top}$.
This possibility and the value of $h_{\rm top}$ depend on various SM parameters.
With current uncertainties (the main ones being on the top quark mass $M_t$ and on the strong coupling $\alpha_3$)
the instability happens within present $\pm2.5\sigma$ uncertainties, and
$h_{\rm top} \sim 10^{11}\GeV \ll \bp$ for current best-fit values.
However $h_{\rm top} $ reaches the Planck scale and the instability disappears for
larger $\alpha_3$ and/or lower $M_t$.
Quantitatively, $h_{\rm top}$ roughly increases by one order of magnitude if $M_t$ decreases by 1 GeV,
a shift indicated by the most recent measurement~\cite{2302.01967}.
Here and in the following we assume numerical values appropriate for $h_{\rm top}\sim 10^{11}\GeV$.

Even if this instability is really present, the consequent vacuum tunnelling rate is negligibly small, 
because suppressed by a non-perturbative factor $e^{-S}$
controlled by the small coupling $\lambda$.
The action of the bounce Higgs field configuration is~\cite{hep-ph/0104016,1604.06090}
\beq S=S_{\rm vacuum} \approx 8\pi^2/3|\lambda| \sim 2000\qquad\hbox{since $-0.01 \circa{<}\lambda<0$.}\eeq
The same conclusion holds for thermal tunnelling at finite temperature:
$S_{\rm thermal} \approx \sfrac{2\pi}{|\lambda|} \sim 600$~\cite{Arnold:1991cv}.

Renewed interest in the possible existence of primordial black holes reopened the issue of vacuum decay
seeded by primordial black holes: can they have existed if the SM potential is unstable?
The minimal allowed values of the exponential suppressions are
\beq S_{\rm vacuum} \sim \ln \frac{h_{\rm top}^4}{H_0^4}\sim 500,\quad
S_{\rm thermal}\sim \ln \frac{\bp^4}{{T}_0^3h_{\rm top}}\sim 200,\quad
S_{\rm BH} \sim \ln  \frac{N_{\rm BH}\bp^3}{h_{\rm top}^3} \sim 50+\ln N_{\rm BH}.\eeq
for vacuum decay (we estimated the space-time volume of our  
past light-cone in units of the bubble volume; the detailed computation can be found e.g.\ in section 6.3 of~\cite{1307.3536}), for decay at finite temperature during the big-bang
(we estimated the space-time volume of the current horizon at temperature $T \sim h_{\rm top}$~\cite{Arnold:1991cv};
a detailed computation can be found e.g.\ in section~3 of~\cite{1608.02555}), and
for vacuum decay seeded by $N_{\rm BH}$ primordial black holes
(we estimated the Hawking evaporation time as $ \sim \bp^2/T_{\rm BH}^3$ setting $T_{\rm BH}\sim h_{\rm top}$
so that the result does not depend on black hole parameters;
a detailed computation can be found e.g.\ in section 3.2 of~\cite{1601.02152}).
We denoted as $\bp = M_{\rm Pl}/\sqrt{8\pi}$ the reduced Planck mass, 
as $H_0 \sim T_0^2/\bp$ the current Hubble rate,
as $T_0$ the current temperature.

\smallskip

Many recent works claim that primordial black holes would have triggered SM vacuum decay before evaporating
because the tunnelling action becomes small $S_{\rm BH} \sim h_{\rm top} R_{\rm BH}$ 
for black holes with Schwarzschild radius $R_{\rm BH}=2GM$ smaller than
the bounce size $\circa{<} 1/h_{\rm top}$,
and thereby with masses $M \circa{<} \bp^2/h_{\rm top}$~\cite{1401.0017,1501.04937,1503.07331,Iliesiu:2015lka,1601.02152,1606.04018,1704.05399,1706.01364}.
This claim has two problems (for related discussions see also~\cite{1706.04523,1708.02138,2111.08017,2209.00639}).
First, these computation have been done in the Euclidean, and thereby they would apply to black holes in
the artificial Hartle-Hawking thermal bath.
Second, the consequent Higgs thermal mass was neglected:
we will show that it keeps Hartle-Hawking vacuum decay 
non-perturbatively suppressed by $1/|\lambda|$ and thereby negligibly slow.

\smallskip

In section~\ref{gen} we reach the same conclusion more in general, covering also 
the physically interesting case of a black hole in empty space (`Unruh vacuum').
This general argument is specialised in section~\ref{bigBH} to large cold black holes, and in section~\ref{sec:HH}
to small hot black holes.
Conclusions are given in section~\ref{concl}.

\smallskip



\section{Argument based on scale invariance}\label{gen}
At first sight, the vacuum decay rate seeded by black holes
depends on three mass scales: 
the black hole mass $M$, the Planck mass $\bp$, and
the instability scale of the scalar potential $h_{\rm top}$.

\smallskip

Since the Higgs instability scale is sub-Planckian,
$h_{\rm top}/ \bp \ll 1$, we can work at leading order in this small parameter.
In this limit the back-reaction of the scalar field on the black hole background is negligible,
Einstein gravity presumably applies, so that the theory is described by the tree-level action 
\beq \label{eq:SSM}
S = \int d^4x \sqrt{|\det g|} \left[-\frac{\bp^2+\xi h^2}{2} R + \frac12 g^{\mu\nu}(\partial_\mu h)(\partial_\nu h) - V(h)+\cdots\right] \eeq
where $\cdots$ denotes the rest of the SM Lagrangian.
We assume that $|\xi| \ll \bp^2/h_{\rm top}^2$, so that the Newton constant is only mildly 
affected by the Higgs.
The limit $h_{\rm top}/ \bp \ll 1$ simplifies the problem:  transitions where the black hole significantly changes mass or even disappears~\cite{1401.0017}
do not happen in this limit.
Vacuum decay can be computed in the fixed  Schwarzschild metric
\beq \label{eq:Schw}
ds^2 =g_{\mu\nu}dx^\mu dx^\nu=   A(r)\,  d t^2- \left[ \frac{ dr^2}{A(r)}+r^2  (d\theta^2+\sin^2\theta\,d\varphi^2)\right] ,\qquad A(r) = 1-  \frac{R_{\rm BH}}{r}\eeq
that depends on the Schwarzschild radius $R_{\rm BH}=2GM$ rather than on $G$ and $M$ separately.
So the problem only depends on two scales: $R_{\rm BH}$
and the instability scale $h_{\rm top}$.

\medskip

%

We show that  the rate of SM vacuum decay seeded by black holes
is non-perturbatively suppressed (and thereby negligibly slow)
by exploiting a special property of the SM Higgs potential: approximate scale invariance. 
Since $h_{\rm top} \gg M_h$ the SM potential can be approximated as $V(h) = \frac14 \lambda h^4$ by neglecting the Higgs mass term.
The whole SM is approximatively scale-invariant, as the RG running of $\lambda<0$ is slow,
being a one-loop perturbative effect:
\beq \lambda(h)\approx -\frac{0.15}{(4\pi)^2} \ln \frac{h^2}{e^{1/2}h_{\rm top}^2}  .\eeq
The numerical value is appropriate for $h_{\rm top}\sim 10^{11}\GeV$.
Thereby the vacuum decay rate cannot depend on the only scale, the Schwarzschild radius $R_{\rm BH}$.
Indeed, by rescaling to dimensionless coordinates $\tilde{x}_\mu = {x}_\mu /R_{\rm BH}$
and the field $h(x)$ to a dimensionless
$\tilde{h}(\tilde{x})= h(x) R_{\rm BH}\sqrt{|\lambda|} $
and inserting $R=0$,\footnote{Extremisation of the total action $S[g,h] = S_0[g] + S_1[g,h] + \cdots$ 
where $h = h_0 + \cdots $ is a small perturbative correction to the dominant gravitational background
$g= g_0 + g_1 + \cdots$ works in the naive way,
 $\min_{g,h}S[g,h] =S_0[g_0] + S_1[g_0,h_0]+\cdots$.}
the action of eq.\eq{SSM} becomes
\beq\label{eq:Srescaled}
S= \frac{1}{|\lambda|} \int d\tilde{t}\int
d\tilde{r} ~4\pi \tilde{r}^2
\left[
\frac{1}{2A}\bigg(
\frac{\partial \tilde{h}}{\partial\tilde{t}}
\bigg)^2 - \frac{A}{2}\bigg(
\frac{\partial \tilde{h}}{\partial \tilde{r}}
\bigg)^2   + \frac{\tilde{h}^4}{4}  + \cdots
\right], \qquad A = 1 - \frac{1}{ \tilde{r}}.
\eeq
where  $\cdots$ again denotes the rest of the SM Lagrangian. 
Scale invariance is slightly broken by quantum effects, and this can be accounted by renormalizing
couplings like $\lambda$ around the relevant energy scale of the problem.
For simplicity we specialised the action to spherical  configurations, making factors of $A$ explicit.
This shows that the decay rate of the Hartle-Hawking or Unruh vacuum resulting from the above action 
is non-perturbatively suppressed as
$ S_{\rm BH}={\cal O}( 1)2\pi/|\lambda| $, unless the ${\cal O}(1)$ factor happens to be zero.

\medskip

The non-minimal coupling $\xi$ of the Higgs to gravity does not enter at this leading order.
This is non-trivial, as the $\xi$ term mixes the Higgs with the graviton:
$h$ affects the Newton constant and thereby the gravitational energy,
so apparently $h$ modifies the background when $\xi\neq0$.
This apparent effect can be removed by the Weyl field redefinition 
\beq \label{eq:Weyl} g_{\mu \nu}=g_{\mu\nu}^{\rm E} /(1+8\pi G \xi h^2)\eeq
that brings the action to the Einstein frame, where no $\xi$ term is present
and the Higgs action is only modified by adding Planck-suppressed extra terms. 
They can be neglected because the SM Higgs potential
$V$ does not contain a large constant term (unlike in an inflationary background).
Explicitly, the Einstein-frame Higgs potential in terms of the re-canonicalised Einstein-frame Higgs $h_{\rm E}$ is 
$V_{\rm E} \simeq \lambda h^4_{\rm E}/4 - 3\lambda \xi (\xi+1/3)h^6_{\rm E}/\bp^2+{\cal O}(h^8/\bp^4)$,
and $V = \frac14\lambda (\sqrt{6}\bP)^4 \sinh^4{h_{\rm E}}/{\sqrt{6}\bP}$ in the conformal limit $\xi=-1/6$.

In the next sections we discuss  more concretely how the non-perturbative suppression 
with a non-zero  ${\cal O}(1)$ factor arises in the SM.
While we are not able of computing the precise value of $S_{\rm BH} \sim S_{\rm thermal}$,
this is enough to show that black holes negligibly source SM vacuum decay.

%
%

\begin{figure}[t]
$$\includegraphics[width=0.45\textwidth]{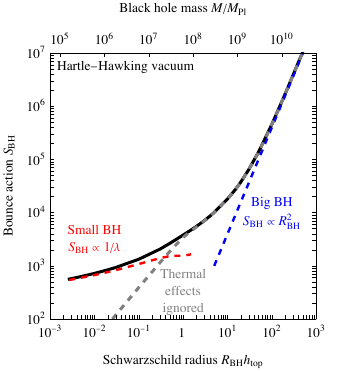}\qquad
\includegraphics[width=0.45\textwidth]{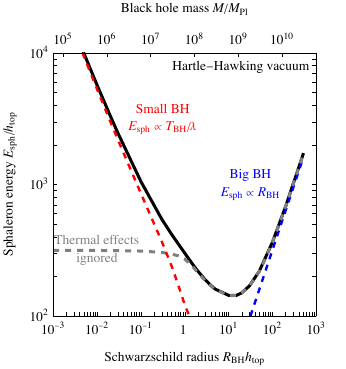}$$
\caption{\it Bounce action $S_{\rm BH} = E_{\rm sph}/T_{\rm BH}$ (left) and
equivalent sphaleron energy $E_{\rm sph}$ (right) 
as function of the Schwarzschild radius $R_{\rm BH}$ in units of the inverse instability scale $h_{\rm top}$.
Only the black hole mass on the upper axis depends on $h_{\rm top}$, here fixed to $10^{11}\GeV\ll M_{\rm Pl}$.
Black curve: numerical result, assuming the Hartle-Hawking vacuum.
Blue dashed curve: asymptotic limit for big cold black holes, see section~\ref{bigBH}.
Red dashed curve: asymptotic limit for small hot black holes, see section~\ref{sec:HH}.
Gray dashed curve: vacuum decay is fast if thermal-like effects are neglected.
Results do not depend on the non-minimal coupling $\xi$ of Higgs to gravity.
\label{fig:SHH}}
\end{figure}

\section{Vacuum decay seeded by big black holes}\label{bigBH}
Issues related to Unruh or Hartle-Hawking vacuum
become irrelevant if the Hawking temperature is negligible compared to the instability scale.
This happens for big heavy black holes.

The effect of their gravity simplifies as the bounce is much smaller than the horizon of the big black hole,
so that the fixed black hole metric background can be approximated by its near-horizon limit.
Indeed, the action gap of eq.\eq{Srescaled} (namely, how much the bounce modifies the action of the larger black-hole background)
is dominated near the horizon.
We focus on the bounce centred on the black hole, as a maximal effect is expected in this simpler case.
Switching to coordinates $t_{\rm R}$ and $r_{\rm R}$ defined as
$t=t_{\rm R}$
and $r \simeq R_{\rm BH}+ r_{\rm R}^2/4R_{\rm BH}$ with small $r_{\rm R} \ll R_{\rm BH}$ near the horizon,
the Schwarzschild metric becomes
\beq  \label{eq:Rindler} ds^2 = - \frac{r_{\rm R}^2}{U}  d(gt_{\rm R})^2+ 
U\,dr_{\rm R}^2 + R_{\rm BH}^2 U^2(d\theta^2+\sin^2\theta\,d\varphi^2)
\qquad \hbox{where}\qquad U = 1+ \frac{r_{\rm R}^2}{4R^2_{\rm BH}}\simeq 1\eeq
and  $g=GM/R_{\rm BH}^2=1/2R_{\rm BH}$ is the surface gravitational acceleration.
Changing coordinates to
$t_{\rm M} = r_{\rm R} \sinh gt_{\rm R}$ and $r_{\rm M} = r_{\rm R} \cosh gt_{\rm R}$
shows that the $t_{\rm R}, r_{\rm R}$ part of eq.\eq{Rindler} is Minkowski in $d=1+1$ dimensions
written in Rindler coordinates.
Rindler coordinates are tied to observers with constant radial acceleration $g$,
and are the Minkowski analogous of Euclidean spherical coordinates. 
Taking also the angular part of eq.\eq{Rindler}  into account, the metric differs from $d=3+1$ Minkowski.
Nevertheless the angular part factorises when considering the
action  of a static (i.e.\ $t_{\rm R}$-independent)
spherical Higgs configuration, 
making its action large and thereby the  tunnelling rate negligible for large $R_{\rm BH}$:
\beq \label{eq:SR}
S_{\rm BH}= 4\pi R_{\rm BH}^2 \int d(g t_R) dr_{\rm R}  r_{\rm R} U \bigg[ \frac{1}{2} \left(\frac{\partial h}{\partial r_{\rm R}}\right)^2
+ U V\bigg] .\eeq
As a result, in the limit $U=1$
the classical Higgs equation of motion that describes the static unstable scalar `hair' around the black hole 
has the same form as in $d=1+1$
\beq \frac{\partial^2 h}{\partial r_R^2} +\left(3-\frac{2}{U}\right) \frac{1}{r_R} \frac{\partial h}{\partial r_R} - U V'=0.\eeq
Being static, the solution $h(r_{\rm R})$ appears to describe the sphaleron of a {\em thermal}-like vacuum decay.
Switching to $t_{\rm M}, r_{\rm M}$ coordinates, $h(\sqrt{r_{\rm M}^2 - t_{\rm M}^2})$
appears to describe the dynamical bounce of {\em quantum} vacuum tunnelling. 
The same phenomenon looks a thermal sphaleron or a quantum bounce in different coordinates because
an accelerated observer perceives a thermal bath at the Unruh temperature $T_{\rm BH}=g/2\pi$.
See~\cite{2105.09331,2209.00639} for related discussions.

\smallskip

The quantum-like description can be computed by continuing $t_{\rm M}=-i\tau_{\rm M}$:
eq.\eq{SR} means that the action $S_{\rm BH}$ for vacuum decay seeded by a large black hole
is proportional to the action $S_{2}$ of flat-space vacuum decay in $d=2$ dimensions.
The gravitational acceleration of the black hole $g$ disappeared in the frame of free-falling observers.

\smallskip

The thermal-like description can be computed by continuing $t_{\rm R}=-i\tau_{\rm R}$
so that $g\tau_{\rm R}$ is an angle with $2\pi$ periodicity.
The action of a static configuration is the time integral of its energy
\beq S_{\rm BH} = \int dt_R E_{\rm sph} =\frac{E_{\rm sph}}{T_{\rm BH}}.
 \eeq
The two continuations describe the same $d=2$ flat Euclidean space: 
$r_{\rm M},\tau_{\rm M}$ are Cartesian coordinates,
$r_{\rm R},g\tau_{\rm R}$ are polar coordinates.
 
The blue curves in fig.\fig{SHH} show the values of $S_{\rm BH}$ and of $E_{\rm sph}$ in the large-black hole limit.
As expected, they asymptotically equal the full result
computed 
by numerically solving the static Higgs equation of motion in a fixed black-hole background
in Schwarzschild coordinates,
\beq h''+h' \left [ \frac{2}{r}- \frac{2GM}{r(r-2GM)}\right]=\frac{V'}{1-2GM/r} ,\eeq
with sphaleron boundary conditions, 
$h=0$ at $r\to\infty$ and $h'= r V'$ at $r=2GM$ on the horizon.\footnote{The figure also covers the 
opposite limit of small black holes. As discussed in the next section, we considered the Hartle-Hawking vacuum
and included in the Higgs potential $V$ a thermal effect, approximating it as
a constant mass term at temperature $T_{\rm BH}$.}
In view of the $R_{\rm BH}^2$ angular factor in eq.\eq{SR},
a black hole makes vacuum decay much slower than in flat space (unlike what happens in $d=1+1$~\cite{2105.09331}).
Different bounces centred away from the black hole have smaller action and dominate the vacuum decay rate.


%

\section{Vacuum decay seeded by small black holes}\label{sec:HH}
We next consider black holes with Hawking temperature comparable or larger than the instability scale, so that
the distinction between Unruh or Hartle-Hawking vacuum becomes relevant.

\begin{figure}[t]
$$\includegraphics[width=0.85\textwidth]{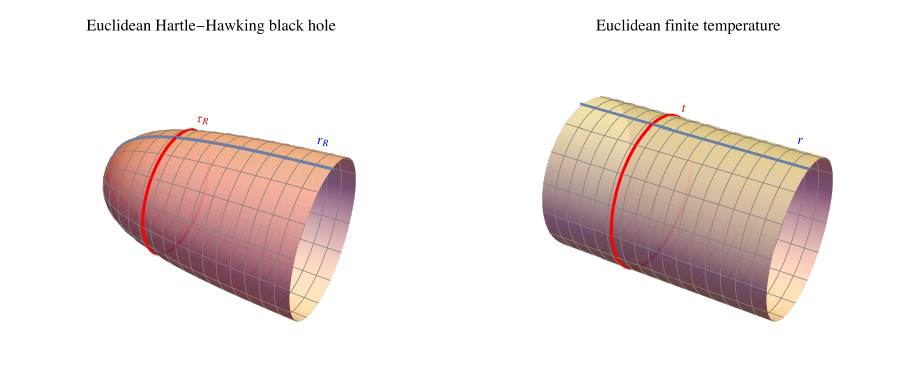}$$
\caption{\it Euclidean Schwarzschild has cigar-like shape. 
Far away from the tip it becomes a cylinder, that describes flat space filled by a thermal bath.
\label{fig:ES}}
\end{figure}

We start from the simpler Hartle-Hawking vacuum, that assumes a black hole in  
equilibrium with a thermal plasma at the Hawking temperature.
This configuration is physically realized only during the big-bang at the specific temperature $T=T_{\rm BH}$.
At the relevant temperatures $T_{\rm BH}\sim h_{\rm top}$ all SM particles are in thermal equilibrium
(despite that some smaller Yukawa couplings such as $y_e$ don't contribute).
This special static system can be computed 
by continuing time $t=-i\tau$ to imaginary time $\tau$ and considering the Euclidean action.
The Schwarzschild metric of eq.\eq{Schw} becomes
\beq -ds^2 = + A(r)\,  d \tau^2+ \frac{ dr^2}{A(r)}+r^2  (d\theta^2+\sin^2\theta\,d\varphi^2) 
\eeq
for $r\ge R_{\rm BH}$ outside the horizon.
This metric is regular at $r=R_{\rm BH}$ avoiding a conical singularity 
if Euclidean time $\tau$ is periodic at the Hawking temperature
\beq \Delta \tau = \frac{1}{T_{\rm BH}} = 8\pi GM = \frac{M}{\bp^2}.\eeq
Fig.\fig{ES}a illustrates the resulting cigar-like Euclidean space-time.
The $\tau$ periodicity holds for any $r$ up to infinity, 
meaning that the corresponding Minkowski space-time is 
filled up to infinity by a thermal bath at the Hawking temperature.
This feature resembles flat space filled by a thermal bath at temperature $T=T_{\rm BH}$, 
where physical quantities such as thermal tunnelling  are
conveniently computed by analytically continuing to an Euclidean cylinder (fig.\fig{ES}b)
with periodicity $1/T$  for Euclidean time $\tau$.
Deviation from thermality encoded in the thinning of the cigar corresponds to the gravitational red-shift  $T(r)=T_{\rm BH}/ \sqrt{A(r)}$
experienced by a local observer at rest in the $r,t$ Schwarzschild coordinates.\footnote{This factor diverges at the horizon, 
 and geometrically corresponds to the cigar radius vanishing at its tip  (red curve in fig.\fig{ES}a).
Some components of the Riemann tensor $R_{\mu\nu\rho\sigma}$ diverge at $r\to 2GM$, 
but this is only a coordinate singularity, as confirmed
by the regularity of the Kretschmann invariant 
$K=R_{\mu\nu\rho\sigma}R^{\mu\nu\rho\sigma} = 48 G^2 M^2/r^6$.
The space is smooth everywhere, so that physical quantities such as $\med{h^2}$ or $\med{T_{\mu\nu}}$ are finite at the horizon~\cite{Candelas:1980zt,book}.
\footnoten{Avoiding a conical singularity is not possible for a dS-S space with two horizons and different temperatures.
The physical interpretation of this could just be that instantons have trouble?
The same issue reappears adding the contribution of the scalar field (I presume, see Gregory), but we bypass the issue
assuming that the mass scale in the scalar potential is very sub-Planckian, so that at leading order in $1/\bp$
we can neglect this effect and compute on a fixed BH background. (to be verified).}}

 \medskip
 
 The limit $R_{\rm sph}\ll R_{\rm BH}$ considered in the previous section~\ref{bigBH}
 corresponds to a bounce so small that the tip of the cigar could be approximated as flat,
 introducing appropriate coordinates.

%
%
A large suppression of $S_{\rm BH}$ has been claimed in the opposite limit
$T_{\rm BH}\gg h_{\rm top}$ i.e.\ for black holes with small mass $M \ll \bp^2/h_{\rm top}$ so that
$R_{\rm BH}$ supposedly is much smaller than the size of the bounce.
If this limit were reached, the near-horizon region (the tip of the cigar) would be negligible, 
and the computation would simplify to a thermal computation on a cylinder.
We keep the full cigar-like geometry because we will see that the this limit is not reached:
the bounce changes size remaining comparable
to the cigar tip.
The sphaleron constant in $\tau$ and O(3)-symmetric has action
\beq\label{eq:Str}
S_{\rm BH}=\int _0^{1/T_{\rm BH}} d\tau \int 4\pi r^2\, dr\, \left[\frac{\dot h^2}{2A}+ A \frac{h'^2}{2} + V  \right]=\frac{E_{\rm sph}}{\TBH}
\eeq
where the 3-dimensional action of the static `sphaleron' configuration 
has been denoted as $E_{\rm sph}$ because
this time-independent configuration also solves the real-time equation of motion and $E_{\rm sph}$ is its energy.
The action seems to scale as $S_{\rm BH} \sim h_{\rm top}/T_{\rm BH}$, leading to the claim that for $\TBH \gg h_{\rm top} $
the non-perturbative suppression of the decay is lost, $S_{\rm BH}\to 0$.

\smallskip

However, one-loop corrections to the potential are important and change this limiting behaviour
when $V$ is promoted to $V_{\rm eff}$, roughly given by the one-loop potential.
Indeed this computation is similar to a thermal computation:
eq.\eq{Str} is similar to a Boltzmann suppression that would disappear at $ \TBH \circa{>} E_{\rm sph}$, if this limit could be reached.
This is prevented by thermal effects that increase $E_{\rm sph} \propto \TBH$ because
the one-loop potential at finite temperature $T$ contains, in addition to quantum effects, 
thermal corrections dominantly described by a thermal mass $m =\kappa  T$ for the $h$ scalar,
$V_{\rm eff} \approx V + m^2 h^2/2$.
This thermal Higgs mass is especially important in the nearly scale-invariant SM, 
where it dominantly sets the scale of the potential, and where
$\kappa\approx 0.35$ is given by~\cite{hep-ph/9212235,hep-ph/9403219}
\beq \label{eq:ThermalMass}\kappa^2= \frac{g_Y^{ 2}+3g^2_2}{16} +\frac{y_t^2}{4} +\frac{\lambda}{2} 
+\cdots .\eeq
The numerical value of $\kappa$ is appropriate for $T\sim h_{\rm top}\sim 10^{11}\GeV$,
and its temperature dependence can be seen e.g.\ in fig.~5 of~\cite{1608.02555}.
Notice that $\kappa\sim 1$ thanks to the top Yukawa $y_t$ and to the gauge couplings $g_{Y,2}$:
this completes the general argument of section~\ref{gen}, 
providing a non-vanishing contribution to the unknown ${\cal O}(1)$ factor.

In the SM at finite temperature,
the sphaleron for the potential $V_{\rm eff}$ 
has action $S_{\rm thermal}=6.015 \pi \kappa/|\lambda|$~\cite{Arnold:1991cv}
non-perturbatively suppressed by the small coupling $\lambda$.
In view of the order-unity thermal mass $\kappa T$, there is no high-temperature regime 
$T\gg h_{\rm top}$ with fast SM thermal decay.
Thermal decay in the SM with $M_h\approx 125\GeV$ remains slower than the Hubble rate at any $T$ 
(see~\cite{1608.02555} for a recent discussion).

\smallskip

The same generic phenomenon happens around a black hole:
the size $R_{\rm sph}$ of the sphaleron remains comparable to $R_{\rm BH}$, and
its action remains non-perturbatively suppressed.
Details differ, and depend on the black hole vacuum. 

Two thermal baths are present in the Hartle-Hawking vacuum:
the outgoing Hawking radiation emitted by the black hole (with flux that scales as $1/r^2$ at large $r$),
and the artificial bath that keeps the system stable.
As a result the Higgs and all SM fields undergo fluctuations, computed in~\cite{Candelas:1980zt} from their propagators,
and roughly approximated for a massless scalar
as $\med{h^2} \sim \TBH^2 +  \TBH^2\,{\cal O}(R_{\rm BH}/r)^2 $.
At $ r \sim R_{\rm BH}$ red-shift modifies the precise dependence on $r$, giving a finite result on the regular horizon~\cite{Candelas:1980zt}.
These field fluctuations induce in the effective action a thermal-like Higgs mass proportional to its couplings\footnote{The other UV-dominated corrections
to the scalar potential associated to the curved geometry discussed in~\cite{1804.02020}
are not present because our black hole background is Ricci-flat ($R_{\mu\nu}=0$ so $R=0$) everywhere (it does not include the source mass $M$ at $r=0$).
See also~\cite{0909.3048,1303.6620} for the one-loop potential in a Schwarzschild background.}

\beq \label{eq:m12} m^2 =  \kappa^2 \TBH^2 [ 1 +  {\cal O}(R_{\rm BH}/r)^2].\eeq
The first effect is mildly dominant because sphalerons have radius a few times larger the horizon,
$R_{\rm sph}\sim 1/m =  (4\pi/\kappa)R_{\rm BH}$ independently of $\lambda$,
making the cylinder approximation reasonable.
To see the qualitative impact of this thermal-like effect, 
we thereby approximate it as a thermal mass $m = \kappa T_{\rm BH}$.
Fig.\fig{SHH}a shows the result: $S_{\rm BH}$ remains large in the small-black hole limit (black curve),
strongly deviating from previous computations that neglected this effect (gray dashed curve),
and becoming similar to the vacuum-decay action (red dashed curve).
Fig.\fig{SHH}b similarly shows that $E_{\rm sph}$ grows as $T_{\rm BH}$.
This approximation might be accurate up to  $\kappa/4\pi \sim 10\%$ or maybe miss ${\cal O}(1)$ factors: 
in any case it makes clear that the vacuum decay rate is negligible.



\medskip

Only the second term of eq.\eq{m12} is present in the Unruh vacuum.
Altought this is not computable in the Euclidean,
the same balance between field fluctuations (that favour vacuum decay)
and thermal-like mass (that suppresses vacuum decay) keeps vacuum decay non-perturbatively suppressed.
At a fixed position $r$, fluctuations up to 
$h > h_{\rm top}(r) = m(r)/\sqrt{|\lambda|}$ above the potential barrier
have been estimated to occur with probability exponentially suppressed by
$S_{\rm BH}^{\rm U}\sim h_{\rm top}^2(r)/2\med{h^2(r)}\sim\kappa^2/|\lambda|$~\cite{1708.02138,2105.09331,2111.08017},
by assuming Gaussian fluctuations of a massless scalar.
A refined estimate that avoids these assumptions can be performed simplifying the geometry, and
viewing a small black hole in the Unruh vacuum as a small region with size $R_{\rm BH}$ at temperature
$T_{\rm BH}=1/4\pi R_{\rm BH}$.
Its vacuum decay rate is presumably dominated by a sphaleron with roughly constant 
$h =h_{\rm top}$ within the small region,
so that $S_{\rm BH}^{\rm U}=E_{\rm sph}^{\rm U}/T_{\rm BH} \sim (4\pi \kappa)^4/12|\lambda|$.
In both estimates the factor $1/\lambda$ (in agreement with the argument based on approximate scale invariance)
together with $\kappa\sim 1$ implies that vacuum decay is negligibly slow.\footnote{The different claim in~\cite{2005.12808} follows from
an Euclidean continuation done keeping only the second term of eq.\eq{m12}: 
but the Euclidean computation is not applicable, and if it were applicable it would add the first term.
Indeed, technically, \cite{2005.12808} compute the full metric perturbed by the scalar as
\beq \label{eq:dsBHh}
ds^2 = e^{2\delta(r)} A(r) d\tau^2 + dr^2/A(r) + r^2 d\Omega_2^2\qquad\hbox{with}\qquad  A(r) = 1- 2GM(r)/r,\eeq
and get the Euclidean action as $S_{\rm BH}=4\pi G [M^2(\infty)- M^2(R_{\rm BH})]$.
In the relevant limit $h_{\rm top}\ll \bp$ this reduces to our eq.\eq{Str},
$S_{\rm BH}=E_{\rm sph}/T_{\rm BH}$
with $E_{\rm sph}=M(\infty)-M(R_{\rm BH})$. This makes clear that one cannot keep the denominator $1/T_{\rm BH}$
while neglecting the effects of the thermal bath on the effective potential.}

%
%


\medskip

Finally, we consider SM vacuum decay seeded by black holes during the big-bang at generic temperature $T$~\cite{1606.04018,1706.01364}.
This rate is again negligible,
as  thermal effects dominate for $T \gg \TBH $; the Hawking-Hartle result applies at $T=\TBH$; 
and the Unruh result at $T\ll \TBH$. Furthermore this finite-temperature vacuum decay rate has to be compared with the Hubble rate
$H \sim T^2/\bp$, which is parametrically faster than the black hole evaporation rate 
$\Gamma_{\rm BH}\sim M^3/\bp^4$, 
that thereby is not relevant during the big bang.

As discussed around eq.\eq{Srescaled}, all these processes negligibly depend on the non-minimal coupling $\xi$
of Higgs to gravity.\footnote{The different claim in~\cite{1706.01364} is based 
on computing the metric of eq.\eq{dsBHh}  at leading order in $G$, finding that $\xi$ contributes as 
\beq \label{eq:dM} M(\infty) - M(R_{\rm BH}) =2\pi R_{\rm BH}\xi h^2(R_{\rm BH}) + \hbox{($\xi$-independent)}.\eeq
The authors of~\cite{1706.01364} interpret this as $E_{\rm sph}$,
and suggest that the corresponding contribution to the action should come from a York boundary term on the horizon.
We disagree for the following reasons.
1) The York term on the horizon would not depend on $G_h \equiv G/(1+8\pi G \xi h^2)$, the Newton constant as modified by the $\xi$ term.
2) $M(R_{\rm BH})$ no longer tells the energy without the Higgs sphaleron, as $G_h\neq G$ modifies the horizon position.
Solving the problem in the Einstein frame, and next introducing a $\xi$ term via the Weyl rescaling of eq.\eq{Weyl},
allows to compute the sphaleron energy as $E_{\rm sph}=M(\infty) - M_{h=0}(\infty)$, which is trivially $\xi$-independent.
3) We agree on eq.\eq{dM}, but the expansion at leading order in $G$ breaks down around the horizon.
The Weyl rescaling allows to compute $M(R_{\rm BH})$ beyond the leading order in $G$, confirming that it
is not a measure of the energy without the sphaleron.
\nnn{We recall that our arguments, including the Einstein-frame argument, apply only in the sub-Planckian limit $h_{\rm top}\ll \bp$.
Beyond this limit a $\xi$-dependence presumably arises, and it's unclear 
how to compute it directly in the Jordan frame}.}


\medskip

Before considering the issue closed, one might want to
compute order unity factors in eq.\eq{m12} to be sure that the squared thermal-like Higgs mass is positive.
We do not see how the standard sign can be changed by the more complicated environment
(non-homogeneous background with non-trivial global geometry, red-shift, out-going Hawking radiation) 
given that thermal masses account for the average energy due to scatterings.
In specific theories squared thermal masses are negative, but for well-understood reasons:
the main example is a new scalar $S$ and a potential term $\lambda_{HS}|H|^2 |S|^2$ with  $\lambda_{HS}$ negative and large enough.
In such a case, large temperature induces symmetry breaking~\cite{Weinberg:1974hy},
and thermal-like effects around black holes would induce fast vacuum decay into the vacuum where $S$ acquires a vacuum expectation value.

\medskip

To conclude, some cautionary words about the black hole interior.
We ignored it because, thanks to the  $h_{\rm top}\ll M_{\rm Pl}$ limit, we computed on a fixed black hole background,
so whatever happens in its interior cannot classically affect the exterior.
However,  the black hole information paradox suggests that at quantum level information might get encoded in
small corrections to thermality of the Hawking radiation. These subtle effects are not expected to affect the vacuum decay issue. 
It could be interesting to consider the effects of a black hole interior in the true vacuum, as it would extend
up to near the horizon.

\section{Conclusions}\label{concl}
Motivated by the possible instability of the SM Higgs potential,
we elaborated on the difficult problem of vacuum decay seeded by black holes.
Instead of solving it, in section~\ref{gen}
we cut the Gordian knot:
by exploiting the sub-Planckian nature of the instability and the
approximate scale invariance of the Standard Model around the instability scale, 
we have shown that Higgs vacuum decay triggered by
primordial black holes is negligibly slow because suppressed by a non-perturbative factor containing the small Higgs quartic coupling $\lambda$,
$e^{{\cal O}(2\pi)/\lambda}$.
This result holds irrespectively of the non-minimal coupling of the Higgs to gravity, $\xi R |H|^2$.

\smallskip

We motivated differences with respect to different claims.
We agree with some recent works that already argued that vacuum decay seeded by black holes is negligibly slow:
Kohri and Marsui~\cite{1708.02138} 
performed an approximated computation using real-time stochastic fluctuations;
Shkerin and Sibiryakov \cite{2105.09331,2111.08017} developed techniques to compute the vacuum decay rate and
computed a 2-dimensional model with features similar to 4-dimensional models.

\smallskip

In section~\ref{bigBH} we considered the simpler limit of vacuum decay seeded by big black holes,
finding that 4-dimensional effects make near-horizon vacuum decay even slower than in flat space
(blue curve in fig.\fig{SHH}).
We discussed how the same vacuum decay can look thermal-like (static sphaleron)
or quantum-like (dynamic bounce) to different observers with relative acceleration.

\smallskip

In section~\ref{bigBH} we considered the more difficult case of vacuum decay seeded by small black holes.
Black holes in the artificial Hartle-Hawking vacuum (i.e.\ surrounded by a thermal bath that keeps the system static)
can be computed via Euclidean analytic computation, finding the same phenomenon that happens at finite temperature:
the non-perturbative suppression persists  at temperature much above the instability scale (red curve in fig.\fig{SHH})
because higher temperatures are counter-balanced by a bigger thermal mass.
This is a order one effect in the Standard Model.
Black holes in the physical Unruh vacuum cannot be computed via Euclidean continuation,
and we estimated its negligible rate.

\medskip

Finally, we mention a more general issue: are primordial black holes compatible with a generic landscape
of deeper vacua separated by Planck-scale potential barriers?
As black holes evaporate up to Planck-scale temperatures, vacuum decay could be enhanced for
heavy scalars with thermal mass smaller than the Higgs (as long as thermal equilibrium still holds),
or in theories with negative squared thermal masses.



\small\frenchspacing
\paragraph{Acknowledgments}
We thank A. Shkerin, S. Sibiryakov, D. Teresi, N. Tetradis for discussions.


\footnotesize
\begin{thebibliography}{nnn}\bibitem{Hiscock:1987hn}
\article{W.A. Hiscock}{Phys.Rev.D}{35}{1161}{1987}
{\href{https://doi.org/10.1103/PhysRevD.35.1161}{Can black holes nucleate vacuum phase transitions?}}.


\bibitem{Arnold:1989cq}
\article{P.B. Arnold}{Nucl.Phys.B}{346}{160}{1990}
{\href{https://doi.org/10.1016/0550-3213(90)90243-7}{Gravity and false vacuum decay rates: O(3) solutions}}.


\bibitem{Berezin:1990qs}
\article{V.A. Berezin, V.A. Kuzmin, I.I. Tkachev}{Phys.Rev.D}{43}{3112}{1991}
{\href{https://doi.org/10.1103/PhysRevD.43.R3112}{Black holes initiate false vacuum decay}}.


\bibitem{Samuel:1992wt}
\article{D.A. Samuel, W.A. Hiscock}{Phys.Rev.D}{45}{4411}{1992}
{\href{https://doi.org/10.1103/PhysRevD.45.4411}{Gravitationally compact objects as nucleation sites for first order vacuum phase transitions}}.


\bibitem{Hartle:1976tp}
\article{J.B. Hartle, S.W. Hawking}{Phys.Rev.D}{13}{2188}{1976}
{\href{https://doi.org/10.1103/PhysRevD.13.2188}{Path Integral Derivation of Black Hole Radiance}}.


\bibitem{Unruh:1976db}
\article{W.G. Unruh}{Phys.Rev.D}{14}{870}{1976}
{\href{https://doi.org/10.1103/PhysRevD.14.870}{Notes on black hole evaporation}}.


\bibitem{1309.0530}
\article[1309.0530]{C. Cheung, S. Leichenauer}{Phys.Rev.D}{89}{104035}{2014}
{\href{https://doi.org/10.1103/PhysRevD.89.104035}{Limits on New Physics from Black Holes}}.


\bibitem{1401.0017}
\article[1401.0017]{R. Gregory, I.G. Moss, B. Withers}{JHEP}{03}{081}{2014}
{\href{https://doi.org/10.1007/JHEP03(2014)081}{Black holes as bubble nucleation sites}}.


\bibitem{1501.04937}
\article[1501.04937]{P. Burda, R. Gregory, I. Moss}{Phys.Rev.Lett.}{115}{071303}{2015}
{\href{https://doi.org/10.1103/PhysRevLett.115.071303}{Gravity and the stability of the Higgs vacuum}}.


\bibitem{1503.07331}
\article[1503.07331]{P. Burda, R. Gregory, I. Moss}{JHEP}{08}{114}{2015}
{\href{https://doi.org/10.1007/JHEP08(2015)114}{Vacuum metastability with black holes}}.


\bibitem{Iliesiu:2015lka}
L.V. Iliesiu, ``\href{https://www.pacm.princeton.edu/sites/pacm/files/iliesiu.pdf}{Black Holes Stimulate Higgs Vacuum Decay}'' (thesis).


\bibitem{1601.02152}
\article[1601.02152]{P. Burda, R. Gregory, I. Moss}{JHEP}{06}{025}{2016}
{\href{https://doi.org/10.1007/JHEP06(2016)025}{The fate of the Higgs vacuum}}.


\bibitem{1606.04018}
\article[1606.04018]{N. Tetradis}{JCAP}{09}{036}{2016}
{\href{https://doi.org/10.1088/1475-7516/2016/09/036}{Black holes and Higgs stability}}.


\bibitem{1704.05399}
\article[1704.05399]{D. Gorbunov, D. Levkov, A. Panin}{JCAP}{10}{016}{2017}
{\href{https://doi.org/10.1088/1475-7516/2017/10/016}{Fatal youth of the Universe: black hole threat for the electroweak vacuum during preheating}}.


\bibitem{1706.01364}
\article[1706.01364]{D. Canko, I. Gialamas, G. Jelic-Cizmek, A. Riotto, N. Tetradis}{Eur.Phys.J.C}{78}{328}{2018}
{\href{https://doi.org/10.1140/epjc/s10052-018-5808-y}{On the Catalysis of the Electroweak Vacuum Decay by Black Holes at High Temperature}}.


\bibitem{1706.04523}
\article[1706.04523]{K. Mukaida, M. Yamada}{Phys.Rev.D}{96}{103514}{2017}
{\href{https://doi.org/10.1103/PhysRevD.96.103514}{False Vacuum Decay Catalyzed by Black Holes}}.


\bibitem{1708.02138}
\article[1708.02138]{K. Kohri, H. Matsui}{Phys.Rev.D}{98}{123509}{2018}
{\href{https://doi.org/10.1103/PhysRevD.98.123509}{Electroweak Vacuum Collapse induced by Vacuum Fluctuations of the Higgs Field around Evaporating Black Holes}}.


\bibitem{2005.12808}
\article[2005.12808]{T. Hayashi, K. Kamada, N. Oshita, J. Yokoyama}{JHEP}{08}{088}{2020}
{\href{https://doi.org/10.1007/JHEP08(2020)088}{On catalyzed vacuum decay around a radiating black hole and the crisis of the electroweak vacuum}}.


\bibitem{2105.09331}
\article[2105.09331]{A. Shkerin, S. Sibiryakov}{JHEP}{11}{197}{2021}
{\href{https://doi.org/10.1007/JHEP11(2021)197}{Black hole induced false vacuum decay from first principles}}.


\bibitem{2111.08017}
\heparticle[2111.08017]{A. Shkerin, S. Sibiryakov}{Black hole induced false vacuum decay: The role of greybody factors}.


\bibitem{2209.00639}
\heparticle[2209.00639]{W.-Y. Ai, J.S. Cruz, B. Garbrecht, C. Tamarit}{Instability of bubble expansion at zero temperature}.


\bibitem{1207.7235}
\article[1207.7235]{{\sc CMS} collaboration}{Phys.Lett.B}{716}{30}{2012}
{\href{https://doi.org/10.1016/j.physletb.2012.08.021}{Observation of a New Boson at a Mass of 125 GeV with the CMS Experiment at the LHC}}.


\bibitem{1207.7214}
\article[1207.7214]{{\sc ATLAS} collaboration}{Phys.Lett.B}{716}{1}{2012}
{\href{https://doi.org/10.1016/j.physletb.2012.08.020}{Observation of a new particle in the search for the Standard Model Higgs boson with the ATLAS detector at the LHC}}.


\bibitem{1112.3022}
\article[1112.3022]{J. Elias-Miro, J.R. Espinosa, G.F. Giudice, G. Isidori, A. Riotto, A. Strumia}{Phys.Lett.B}{709}{222}{2012}
{\href{https://doi.org/10.1016/j.physletb.2012.02.013}{Higgs mass implications on the stability of the electroweak vacuum}}.


\bibitem{1307.3536}
\article[1307.3536]{D. Buttazzo, G. Degrassi, P.P. Giardino, G.F. Giudice, F. Sala, A. Salvio, A. Strumia}{JHEP}{12}{089}{2013}
{\href{https://doi.org/10.1007/JHEP12(2013)089}{Investigating the near-criticality of the Higgs boson}}.


\bibitem{2302.01967}
\heparticle[2302.01967]{CMS collaboration}{Measurement of the top quark mass using a profile likelihood approach with the lepton+jets final states in proton-proton collisions at $\sqrt{s}$ = 13 TeV}.


\bibitem{hep-ph/0104016}
\article[hep-ph/0104016]{G. Isidori, G. Ridolfi, A. Strumia}{Nucl.Phys.B}{609}{387}{2001}
{\href{https://doi.org/10.1016/S0550-3213(01)00302-9}{On the metastability of the standard model vacuum}}.


\bibitem{1604.06090}
\article[1604.06090]{A. Andreassen, D. Farhi, W. Frost, M.D. Schwartz}{Phys.Rev.D}{95}{085011}{2017}
{\href{https://doi.org/10.1103/PhysRevD.95.085011}{Precision decay rate calculations in quantum field theory}}.


\bibitem{Arnold:1991cv}
\article{P.~B.~Arnold, S.~Vokos}{Phys. Rev. D}{44}{3620}{1991}{Instability of hot electroweak theory: bounds on $M_h$ and $M_t$}.


\bibitem{1608.02555}
\article[1608.02555]{A. Salvio, A. Strumia, N. Tetradis, A. Urbano}{JHEP}{09}{054}{2016}
{\href{https://doi.org/10.1007/JHEP09(2016)054}{On gravitational and thermal corrections to vacuum decay}}.


\bibitem{Candelas:1980zt}
\article{P. Candelas}{Phys.Rev.D}{21}{2185}{1980}
{\href{https://doi.org/10.1103/PhysRevD.21.2185}{Vacuum Polarization in Schwarzschild Space-Time}}.


\bibitem{book} V. Frolov and I. Novikov, ``Black Hole Physics'', Kluwer Academic (1997).


\bibitem{hep-ph/9212235}
\article[hep-ph/9212235]{P.B. Arnold, O. Espinosa}{Phys.Rev.D}{47}{3546}{1993}
{\href{https://doi.org/10.1103/PhysRevD.47.3546}{The effective potential and first order phase transitions: Beyond leading-order}}.


\bibitem{hep-ph/9403219}
\article[hep-ph/9403219]{Z. Fodor, A. Hebecker}{Nucl.Phys.B}{432}{127}{1994}
{\href{https://doi.org/10.1016/0550-3213(94)90596-7}{Finite temperature effective potential to order $g^4, \lambda^2$ and the electroweak phase transition}}.


\bibitem{1804.02020}
\article[1804.02020]{T. Markkanen, S. Nurmi, A. Rajantie, S. Stopyra}{JHEP}{06}{040}{2018}
{\href{https://doi.org/10.1007/JHEP06(2018)040}{The 1-loop effective potential for the Standard Model in curved spacetime}}.


\bibitem{0909.3048}
\article[0909.3048]{P.O. Kazinski}{Phys.Rev.D}{80}{124020}{2009}
{\href{https://doi.org/10.1103/PhysRevD.80.124020}{One-loop effective potential of the Higgs field on the Schwarzschild background}}.


\bibitem{1303.6620}
\article[1303.6620]{C. Garcia-Recio, L.L. Salcedo}{Class.Quant.Grav.}{30}{097001}{2013}
{\href{https://doi.org/10.1088/0264-9381/30/9/097001}{The perturbative scalar massless propagator in Schwarzschild spacetime}}.


\bibitem{Weinberg:1974hy}
\article{S. Weinberg}{Phys. Rev.}{D9}{3357}{1974}{Gauge and Global Symmetries at High Temperature}.


\end{thebibliography}
\end{document}